\documentclass{elsart}
\usepackage{epsfig}
\usepackage{graphicx}

\begin{document}

\begin{frontmatter}



\title{Studies of Stability and Robustness for Artificial Neural Networks and Boosted Decision Trees}

\author{Hai-Jun Yang$^{a,c,*}$, Byron P. Roe$^a$, Ji Zhu$^b$}
\address{$^a$ Department of Physics, University of Michigan, Ann Arbor, MI 48109, USA}\address{$^b$ Department of Statistics, University of Michigan, Ann Arbor, MI 48109, USA}
\address{$^c$ Los Alamos National Laboratory, Los Alamos, NM 87545, USA}

\corauth[yang]{Corresponding author. E-mail address: yhj@umich.edu}

\begin{abstract}

In this paper, we compare the performance, stability and robustness of
Artificial Neural Networks (ANN) and
Boosted Decision Trees (BDT)
using MiniBooNE Monte Carlo samples.  These methods attempt to
classify events given a number of identification variables.
The BDT algorithm has been
discussed by us in previous publications.
Testing is done in this paper by
smearing and shifting the input variables of testing samples.
Based on these studies, BDT
has better particle identification performance than ANN.
The degradation of the classifications obtained by shifting or
smearing variables of testing results
is smaller for BDT than for ANN.

\end{abstract}


\end{frontmatter}

\section{Introduction}

The boosting algorithm is one of the most powerful learning techniques introduced
during the past decade.
The motivation for the boosting algorithm is to design a procedure that
combines many ``weak'' classifiers (such as decision trees, random forests, ANNs etc.) to achieve
a powerful classifier. One starts with unweighted training events and builds 
a decision tree\cite{nima-boosting2005}.
If a training event is misclassified, then the weight of that event is increased (boosted).
A second tree is built using exactly the same set of training events but with new weights.
Again misclassified events have their weights boosted and the procedure is repeated
several hundred to thousand times until the performance becomes optimal. Each
test event is followed through each tree in turn. If it lands on a signal leaf
it is given a score of 1, otherwise $-1$. The sum of the scores from all trees,
is the final score of the event. A high score means that the event is most likely signal and a
low score that it is most likely background. The major advantages of boosted decision trees are
their stability, their ability to handle large numbers of input variables, and their
use of boosted weights for misclassified events to give these events a better chance
to be correctly classified in succeeding trees.

The Artificial Neural Network (ANN) technique has been widely used in data analysis
of High Energy Physics (HEP) experiments in the last decade. The use of the ANN 
technique usually gives better results than the traditional simple-cut techniques.
Based on our previous studies, Boosted Decision Trees (BDT) with the
Adaboost\cite{adaboost1,adaboost2,adaboost3} or $\epsilon-$Boost\cite{eboost1,eboost2}
algorithms perform better than ANN and some other boosting algorithms
for MiniBooNE particle identification (PID)\cite{nima-boosting2005,nima-boosting2005-2}.
More and more major HEP experiments (ATLAS,BaBar,CDF,D0 etc.)
\cite{atlas1,babar1,babar2,babar3,tevatron1,tevatron2,tool} have begun to 
use boosting algorithms as an 
important tool for data analysis since our first successful application of BDT 
for MiniBooNE PID\cite{nima-boosting2005,nima-boosting2005-2}. 

In this paper we discuss the Boosting method
in the context of MiniBooNE.
For practical applications of data mining algorithms, performance, stability and
robustness are determinants.
We focus on stability and robustness of ANN and BDT with $\epsilon-$Boost
($\epsilon = 0.01$)
by smearing or shifting values of input variables randomly for testing samples.
The results obtained in this paper do not 
represent optimal MiniBooNE PID performance
because we only use 30 arbitrarily selected variables for ANN and BDT training and testing.
BDT with more input variables results in significantly 
better performance.  However, 
ANN will not improve significantly by using more input variables
\cite{nima-boosting2005,nima-boosting2005-2}.
MiniBooNE is a crucial experiment operated at Fermi National Accelerator Laboratory which
is designed to confirm or refute the evidence for $\nu_\mu \rightarrow \nu_e$ oscillations at 
$\Delta m^2 \simeq 1 eV^2$ seen by the LSND experiment\cite{boone,lsnd}.
It will imply new physics beyond the Standard Model of particle physics if the LSND signal  
is confirmed by the MiniBooNE experiment.

\section{Training and Testing Samples}

The training sample has 50000 signal and 100000 background events.
An independent testing sample has 54200 signal and 146600 background events.
Fully oscillated $\nu_e$ charged current quasi-elastic (CCQE) events are
signal; all $\nu_\mu$ and non-CCQE intrinsic $\nu_e$ events are treated
as background. The signature of each event is given by 322 variables\cite{sfitters,rfitters}. 
Thirty out of 322 variables were selected randomly for this study.  
(The selection was by variable name not by the power of the variables.) All 
selected variables are used for ANN and BDT training and testing.

The detailed description of reconstructed variables is available in MiniBooNE
technical notes\cite{sfitters,rfitters}. Here we briefly mention some of 
variables used for this study. The MiniBooNE neutrino detector is filled
with 800 tons of pure mineral oil, which is contained in a spherical tank
of 610-cm inner radius. The detector is divided into two optical isolated
regions. The inner region employs 1280 8-inch photomultiplier tubes (PMTs) to 
measure the charge, time and position of particles produced by neutrino interactions
on nuclei. The outer veto region is instrumented with 240 PMTs to 
tag particles entering or leaving the detector.
The final state charged particles passing through the oil can emit both
prompt Cherenkov and delayed isotropic scintillation photons, which are 
detected in a ratio of about 3:1 for $\beta \sim 1$ particles. 
Some reconstructed variables are listed in the following:
\begin{itemize}
\item fraction of very prompt PMT hits, $-2 ~ns < \Delta T < 3 ~ns$
\item fraction of very late PMT hits, $\Delta T > 15 ~ns$
\item reconstructed track length
\item angle between track direction and neutrino beam direction
\item ratio of scintillation flux to Cherenkov flux
\item time likelihood in the Cherenkov ring region $0.55 <\cos\theta <0.85$
\item time likelihood in $0.4 < \cos \theta < 0.5$ 
\item time likelihood in $0.2 < \cos \theta < 0.4$ 
\item time likelihood in $-0.2 < \cos \theta < 0.2$ 
\item time likelihood in $\cos \theta < -0.2$ 
\item charge likelihood in various $\cos \theta $ regions 
\item reconstructed $\pi^0$ mass
\item angle between two $\gamma$'s  from $\pi^0$ decay
\item distance from the first $\gamma$ conversion point to the tank wall
\item the difference of time likelihoods between electron and $\pi^0$
\item the difference of charge likelihoods between electron and $\pi^0$
\item ratio of measured to predicted charge in $\cos \theta < -0.8$ 
\item ratio of measured to predicted charge in $0.5 < \cos \theta < 0.7$ 
\item for the two $\gamma$ hypothesis, the fraction of  
Cherenkov flux in the lower energy $\gamma$ 
\end{itemize}

The time (charge) likelihood is the likelihood of the time (charge) distribution 
of the PMT hits under the given hypothesis.

\subsection{Training Samples}

We prepared 10 statistically independent training samples.  Each sample has 5000 signal and 
10000 background events selected sequentially from the large training sample.
Both ANN and BDT are trained separately on each of these 
training samples.
For a given testing sample, then, ANN and BDT each have 10 sets of results. 
The mean values and
variance of the 
10 sets of results are calculated to compare the ANN and BDT methods.

\subsection{Testing Samples Set 1 - Smearing Randomly}

In order to study the stability of ANN and BDT on the testing samples, we randomly 
smear the input variables by 1\%, 3\%, 5\%, 8\% and 10\%, respectively. The smearing
formula is written as
$$
 V^j_i = V^j_i \times (1 + sf \times R^j_i)
$$
where $V^j_i$ represents value of $j$-th variable in $i$-th testing event,
$sf$ is the smearing factor (= 0, 0.01, 0.03, 0.05, 0.08, 0.1). 
$R^j_i$ is a random number with a Gaussian distribution; it is different 
for each variable and each event. 

\subsection{Testing Samples Set 2 - Shifting Randomly}

The random shift formula can be written as
$$
 V^j_i = V^j_i \times (1 + sf \times R^j_i)
$$
where $V^j_i$ represents value of $j$-th variable in $i$-th testing event,
$sf$ is the shifting factor (= 0, 0.01, 0.03, 0.05, 0.08, 0.1) and 
$R^j_i$ is a discrete random number with value 1 or $-1$.

\subsection{Testing Samples Set 3 - Shifting Positively}

The values of all training variables are shifted positively. The formula is,
$$
 V^j_i = V^j_i \times (1 + sf)
$$
where $V^j_i$ represents value of $j$-th variable in $i$-th testing event,
$sf$ is the shifting factor (= 0, 0.01, 0.03, 0.05, 0.08, 0.1).

\subsection{Testing Samples Set 4 - Shifting Negatively}

The values of all training variables are shifted negatively. The formula is,
$$
 V^j_i = V^j_i \times (1 - sf)
$$
where $V^j_i$ represents value of $j$-th variable in $i$-th testing event,
$sf$ is the shifting factor (= 0, 0.01, 0.03, 0.05, 0.08, 0.1).

\subsection{Testing Samples Set 5 - Shifting Mix}

Each variable is shifted in one direction for all testing events.  The shift direction
for each variable is determined by a random number with the discrete values of 1 or $-1$..
The formula is written as,
$$
 V^j_i = V^j_i \times (1 + sf \times R^j) 
$$
where $V^j_i$ represents value of $j$-th variable in $i$-th testing event,
$sf$ is the shifting factor (= 0, 0.01, 0.03, 0.05, 0.08, 0.1).

\section{Results}

All ANN and BDT results shown in this paper are from testing samples.

\subsection{Results from original testing samples}

Table 1 lists the signal and background efficiencies for ANN and BDT with
root mean square (RMS) errors and statistical errors for background efficiencies. 
The efficiency ratio is defined as background efficiency from ANN divided by 
that from BDT using the original testing sample (no smearing and shifting) 
and the same signal efficiency. Efficiency ratio values greater than 1 mean
that BDT works better than ANN by suppressing more background events (less background
efficiency) for a given signal efficiency. From Table 1, the efficiency
ratios vary from about 1.12 to 1.75 for signal efficiencies ranging from 
90\% to 30\%. Lower signal efficiencies yield higher ratio values. 
The statistical error of the test background efficiency for ANN is slightly
higher than that for BDT depending on the signal efficiency. 
The variance of 10 test background efficiencies for ANN trained with
10 randomly selected training samples is about $2 \sim 5$ times larger
than that for BDT. This result indicates that BDT training performance is more
stable than ANN training.

\subsection{Results from smeared testing samples}

The background efficiency versus signal efficiency for the smeared testing 
sample set 1 is shown in Figure 1. The top plot shows results from ANN, 
the bottom plot shows
results from BDT. Dots are for the results from the testing sample without smearing,
boxes, triangles, stars, circles and crosses are for results from testing samples with 
1\%, 3\%, 5\%, 8\% and 10\% smearing, respectively.
Both ANN and BDT are quite stable for testing samples 
which are randomly smeared within 5\%, typically within about 5\%-12\% performance decrease
for BDT and 7\% - 16\% decrease for ANN as shown in Figure 1. 
For the 10\% smeared testing sample, however,
the performance of ANN is degraded by 37\% to 62\%; higher signal efficiency
results have larger degradation. The corresponding performance of BDT 
is degraded by 19\% to 57\%. 

The variance (RMS) of background efficiencies based on trials versus
signal efficiency for the 10 different
smeared testing samples is shown in Figure 2.
The variance of background efficiencies from BDT is about $2 \sim 5$ times smaller than
that from ANN as presented in the bottom plot of Figure 3. The variance 
ratios between ANN and 
BDT remain reasonably stable for various testing samples with different smearing factors. 

Figure 3 shows the ratio of background efficiency from ANN and BDT versus signal efficiency 
(top plot)
and the ratio of RMS of background efficiency from ANN and BDT versus signal efficiency 
(bottom plot).
Dots are for results from the testing sample without smearing;
boxes, triangles, stars, circles and crosses are for results from 1\%, 3\%, 5\%,
8\% and 10\% smearing, respectively.
Error bars in the top plot are for RMS errors of ratios which are calculated 
by propagating errors from the  RMS errors from ANN and BDT results. 
The performance of BDT ranges from  12\% to 75\% better than that of ANN, 
depending on the signal efficiency as shown in the top plot of Figure 3. 
The ratio of background efficiency 
from ANN and BDT increases with an increase in the smearing factor. 
For the testing sample with 10\% random smearing, 
the efficiency ratio increases about 15\%. This result indicates that the BDT is more
stable than ANN for this set of specific testing samples with random smearing.

Figure 4 shows the background efficiency versus smearing factor for three
given signal efficiencies 30\%(dots) , 50\%(boxes) and 70\%(triangles).
The top plot of Figure 4 shows results from ANN. The bottom plot of Figure 4 shows
results from BDT. The performance of ANN and BDT degrade modestly with
relatively small smearing factors $> \sim$ 0.03. Larger smearing factors
result in significant performance degradation of ANN and BDT.

\subsection{Results from shifted testing samples}

Figure 5 shows the ratio of background efficiency from ANN and BDT versus signal efficiency (top plot)
and the ratio of variance of background efficiency from ANN and BDT versus signal efficiency (bottom plot)
for the testing samples of set 2 with random shifting.
Figure 6 shows the background efficiency versus smearing factor, for ANN (top plot) and BDT (bottom plot).
Resutls from the randomly shifted testing samples of set 2 are comparable to those from the randomly
smeared testing samples of set 1.
In order to estimate the dependence of ANN and BDT performance on the factor of random shifting,
we vary the shifting factor from 0 to 0.5, as shown in Figure 7. Both ANN and BDT performance degrade
significantly by randomly shifting the input variables with large shifting factor( $>$ 0.1).

Figure 8 shows the ratio of background efficiency from ANN and BDT versus signal efficiency (top plot)
and the ratio of variance of background efficiency from ANN and BDT versus signal efficiency (bottom plot)
for the testing samples of set 3 with overall positive shifting.
Figure 9 shows the background efficiency versus smearing factor, for ANN (top plot) and BDT (bottom 
plot).
The results are reasonably stable for both ANN and BDT versus shifting factor.

Figure 10 shows the ratio of background efficiency from ANN and BDT versus signal efficiency (top plot)
and the ratio of variance of background efficiency from ANN and BDT versus signal efficiency (bottom plot)
for the testing samples of set 4 with overall negative shifting.
Figure 11 shows the background efficiency versus smearing factor, for ANN (top plot) and BDT (bottom 
plot).
The results are remain reasonably stable for both ANN and BDT versus shifting factor.

Figure 12 shows the ratio of background efficiency from ANN and BDT versus signal efficiency (top plot)
and the ratio of variance of background efficiency from ANN and BDT versus signal efficiency (bottom plot)
for the testing samples of set 5 with random positive or negative shifting.
Figure 13 shows the background efficiency versus smearing factor, for ANN (top plot) and BDT (bottom 
plot).

\subsection{Further Validation}

In order to make a cross check, a new set of 30 out of the 322 
particle identification variables were selected and the whole analysis
was redone.  Most results are quite similar to the results obtained in 
Sections 3.1--3.3.  BDT, again, was more stable than ANN.  However, the 
second set of 30 variables overall was less powerful by a  factor of about 2 
than the first set.
Because of this, the variances were dominated more by the random
variations than the variations due to a change in power with smearing
or shifting.  The variances of the second set were only about half
the variances of the first set, but exhibited much more random behavior.

\section{Conclusions}

The performance, stability and robustness of ANN and BDT were compared for particle
identification using the MiniBooNE Monte Carlo samples. 
BDT has better particle identification performance than ANN, even using only 30
PID variables.
The BDT performance relative to that of ANN depends on the signal efficiency.
The variance in  background efficiencies of testing results due to various BDT trainings 
is smaller than those from ANN trainings regardless of
testing samples with or without smearing and shifting.
The performance of both BDT and ANN are degraded 
by smearing and shifting the input variables of
the testing samples. ANN degrades more than BDT depending on 
the signal efficiency based on MiniBooNE Monte Carlo samples.

\section{Acknowledgments}
We wish to express our gratitude to the MiniBooNE collaboration for the excellent
work on the Monte Carlo simulation and the software package for physics analysis.
This work is supported by the Department of Energy and by the 
National Science Foundation of the United States.



\begin{table}
\begin{center}
{\small {
\begin{tabular}{|c|c|c|c|c|} \hline
 Eff\_signal (\%) & Eff\_bkgd\_ANN  (\%) & Eff\_bkgd\_BDT (\%) & Ratio = Eff\_bkgd\_ANN \\ 
                  & $\pm~ \sigma_{RMS}$   & $\pm~ \sigma_{RMS}$  & / Eff\_bkgd\_BDT \\ \hline
 30 &    0.416 $\pm$ 0.045 &    0.238 $\pm$    0.009 &    1.748 $\pm$    0.201  \\ \hline

 35 &    0.512 $\pm$ 0.051 &    0.310 $\pm$    0.008 &    1.655 $\pm$    0.172  \\ \hline

 40 &    0.623 $\pm$ 0.057 &    0.389 $\pm$    0.014 &    1.599 $\pm$    0.157  \\ \hline

 45 &    0.748 $\pm$ 0.063 &    0.481 $\pm$    0.017 &    1.557 $\pm$    0.142  \\ \hline

 50 &    0.885 $\pm$ 0.070 &    0.594 $\pm$    0.027 &    1.489 $\pm$    0.136  \\ \hline

 55 &    1.041 $\pm$ 0.076 &    0.722 $\pm$    0.024 &    1.441 $\pm$    0.116  \\ \hline
 
 60 &    1.227 $\pm$ 0.085 &    0.882 $\pm$    0.023 &    1.391 $\pm$    0.102  \\ \hline

 65 &    1.449 $\pm$ 0.088 &    1.074 $\pm$    0.023 &    1.350 $\pm$    0.088  \\ \hline

 70 &    1.732 $\pm$ 0.094 &    1.315 $\pm$    0.026 &    1.317 $\pm$    0.076  \\ \hline

 75 &    2.095 $\pm$ 0.102 &    1.643 $\pm$    0.036 &    1.276 $\pm$    0.068  \\ \hline

 80 &    2.585 $\pm$ 0.111 &    2.110 $\pm$    0.049 &    1.225 $\pm$    0.060  \\ \hline

 85 &    3.316 $\pm$ 0.124 &    2.842 $\pm$    0.079 &    1.167 $\pm$    0.054  \\ \hline

 90 &    4.618 $\pm$ 0.129 &    4.143 $\pm$    0.113 &    1.115 $\pm$    0.044  \\ \hline

\end{tabular}
}}
\end{center}
\caption{Signal and background efficiencies for ANN and BDT with
RMS errors for background efficiencies. 
The ratio is defined as the background efficiency
from ANN divided by that from BDT using the original testing sample
(no smearing and shifting) and the same signal efficiency.}
\end{table}

\newpage

\begin{figure}
\epsfig{figure=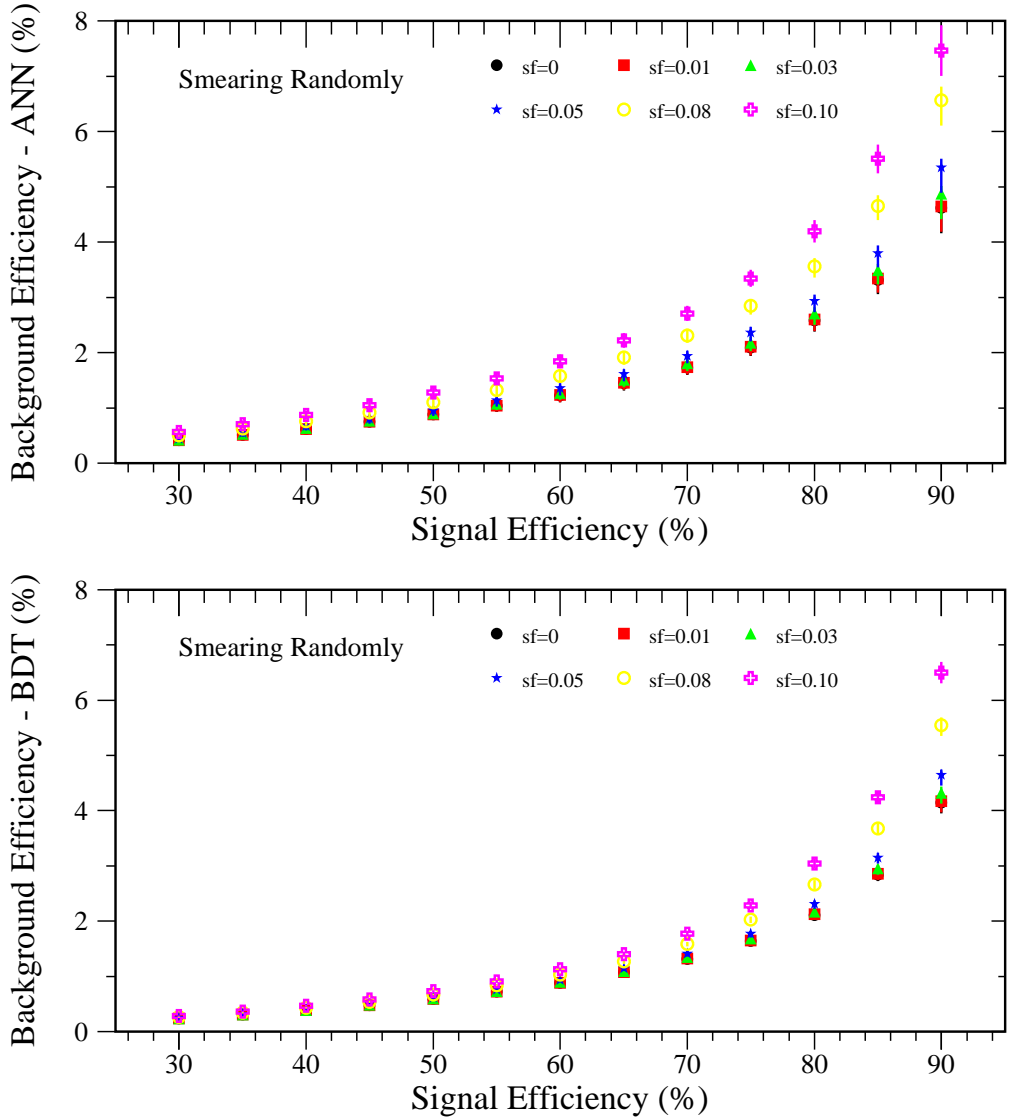,width=14cm}
\caption{Background efficiency versus signal efficiency. The top plot
shows results from ANN with smeared testing samples set 1. 
The bottom plot shows results from BDT.
Dots are for the testing sample without smearing; boxes, triangles, stars,
circles and crosses are for 1\%, 3\%, 5\%, 8\% and 10\% smearing, respectively.}
\end{figure}

\begin{figure}
\epsfig{figure=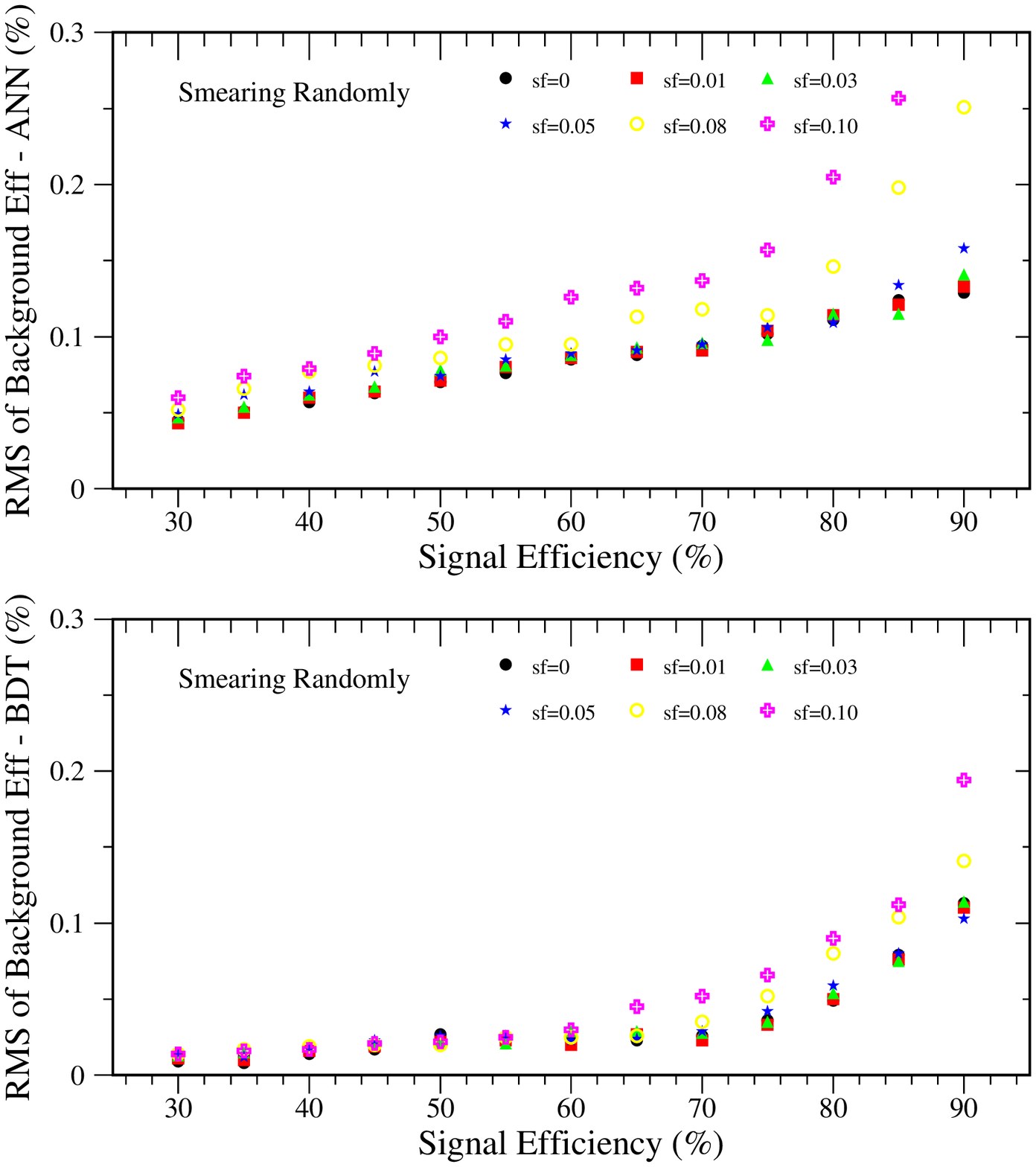,width=14cm}
\caption{Variance of background efficiencies versus signal efficiency. 
The top plot
shows results from ANN with smeared testing samples set 1. 
The bottom plot shows results from BDT.
Dots are for the testing sample without smearing; boxes, triangles, stars,
circles and crosses are for 1\%, 3\%, 5\%, 8\% and 10\% smearing, respectively.}
\end{figure}

\begin{figure}
\epsfig{figure=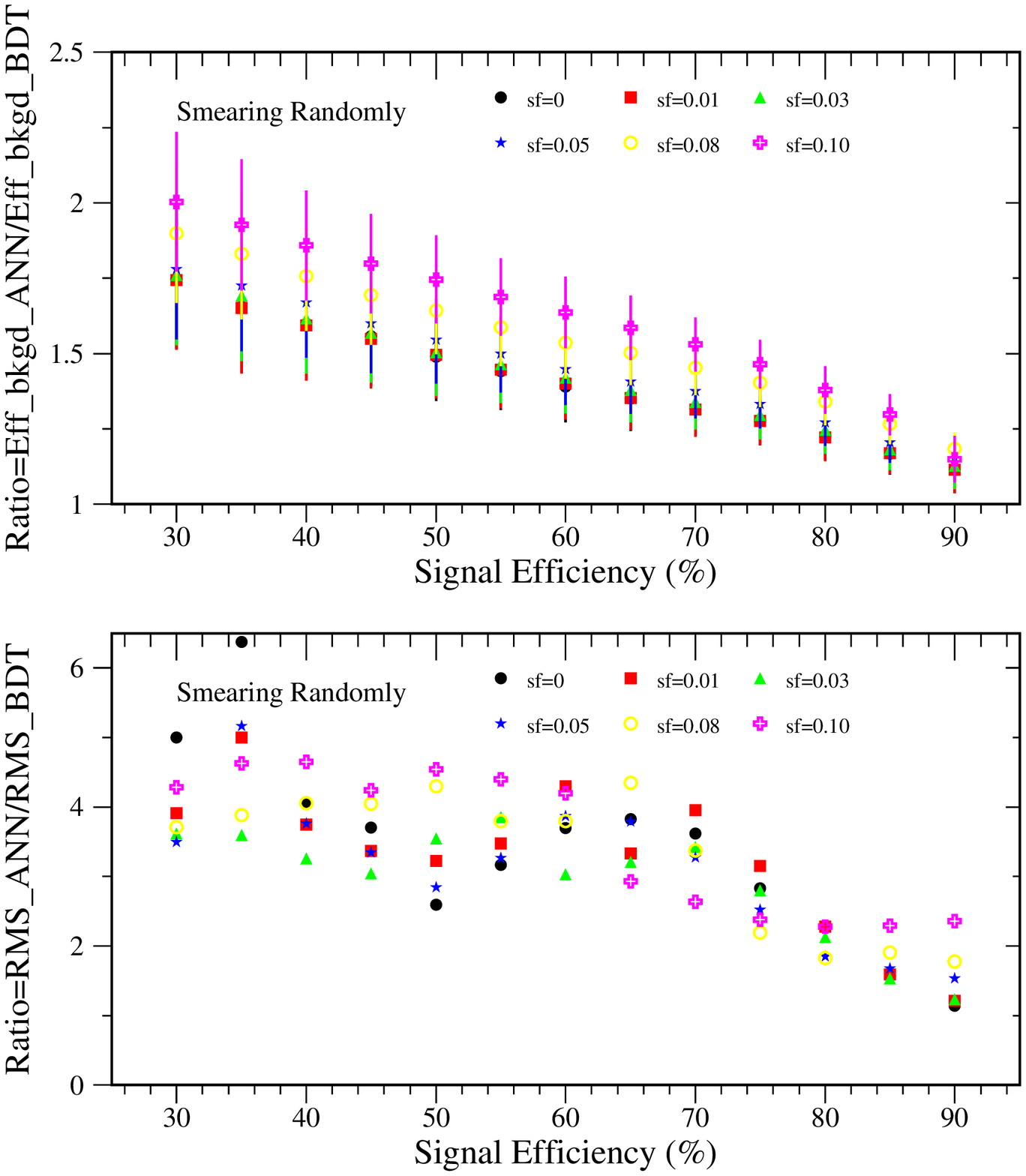,width=14cm}
\caption{Ratio of background efficiency from ANN divided by that from BDT
 versus signal efficiency(top plot) and ratio of variance from ANN divided by that
from BDT versus signal efficiency(bottom plot) with smeared testing samples set 1.
Dots are for the testing sample without smearing; boxes, triangles, stars,
circles and crosses are for 1\%, 3\%, 5\%, 8\% and 10\% smearing, respectively.}
\end{figure}

\begin{figure}
\epsfig{figure=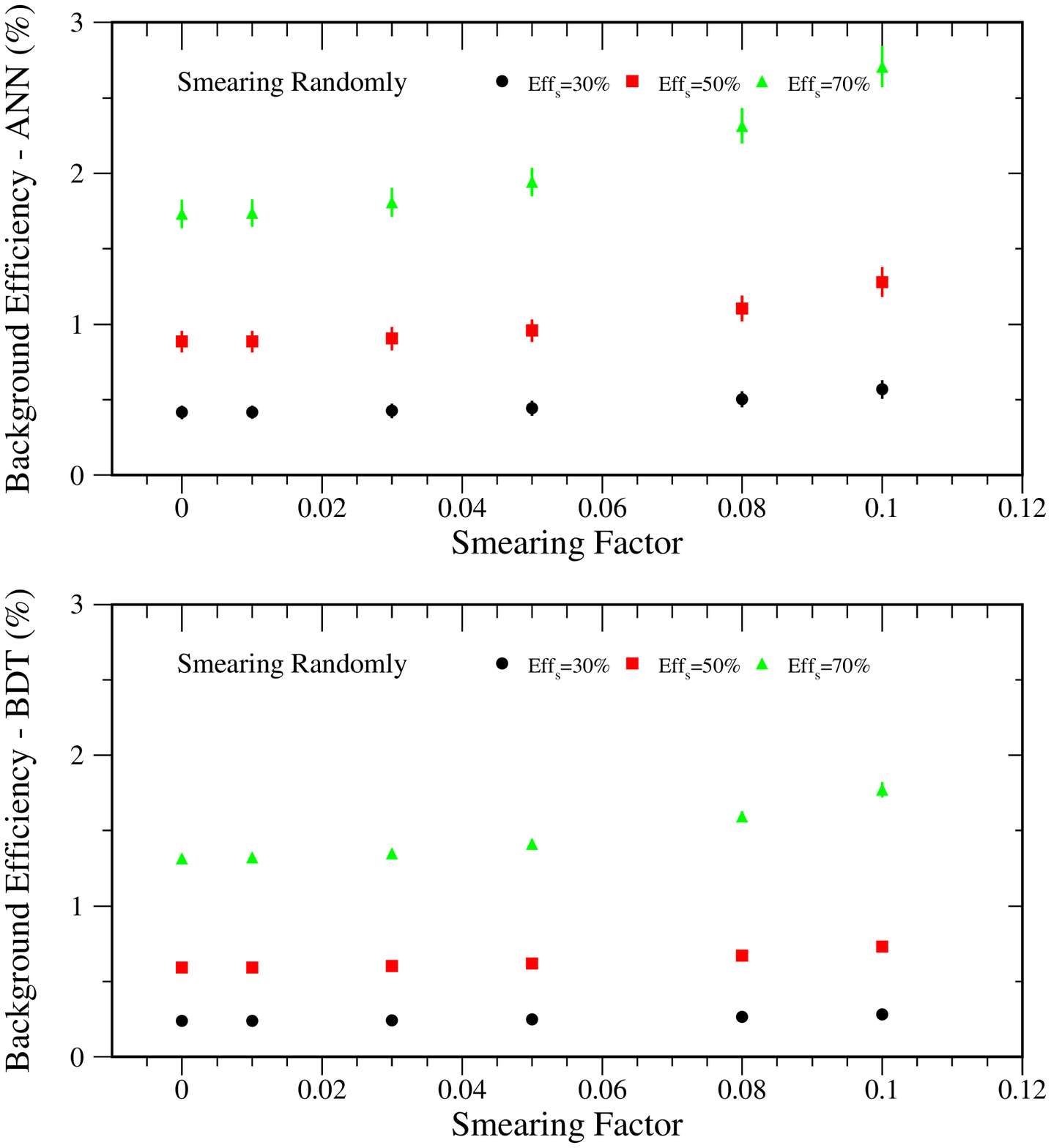,width=14cm}
\caption{Background efficiency versus smearing factor. The top plot
shows results from ANN with smeared testing samples set 1. 
The bottom plot shows results from BDT. Dots, boxes and triangles are for
30\%, 50\% and 70\% signal efficiency, respectively.}
\end{figure}


\begin{figure}
\epsfig{figure=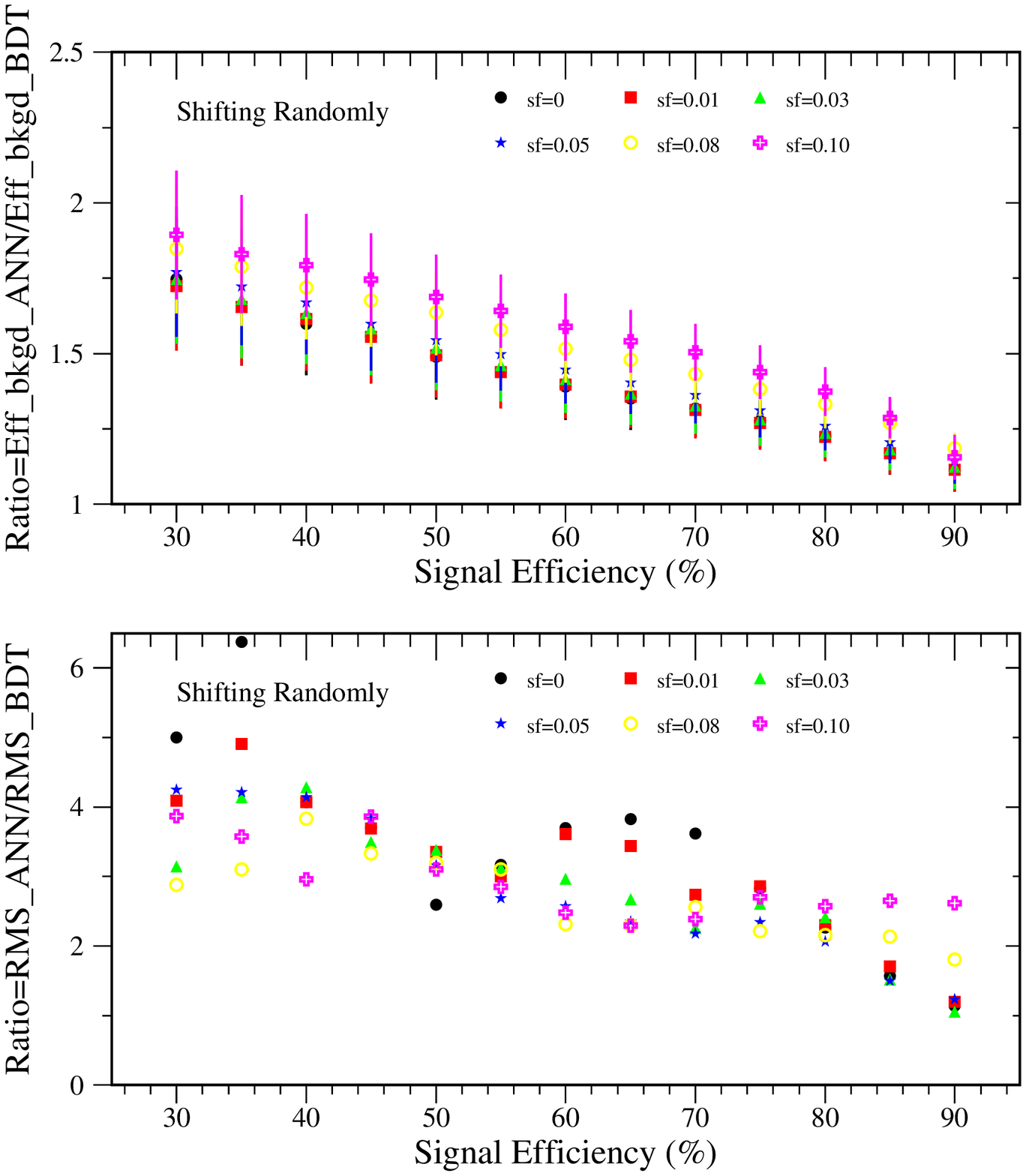,width=14cm}
\caption{Ratio of background efficiency from ANN divided by that from BDT
 versus signal efficiency(top plot) and ratio of variance from ANN divided by that
from BDT versus signal efficiency(bottom plot) with testing samples set 2.
Dots are for the testing sample without shifting; boxes, triangles, stars,
circles and crosses are for 1\%, 3\%, 5\%, 8\% and 10\% shifting, respectively.}
\end{figure}

\begin{figure}
\epsfig{figure=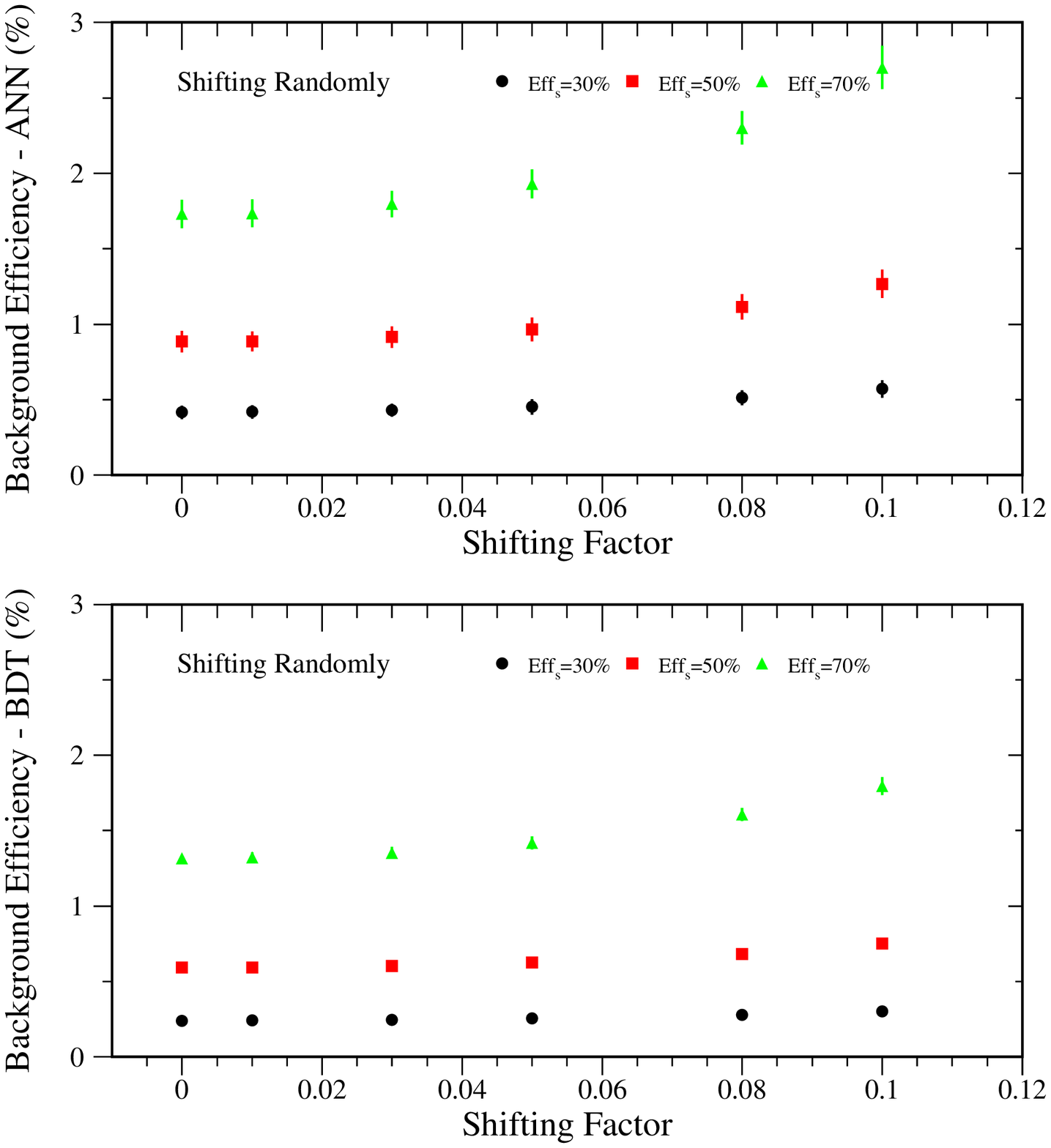,width=14cm}
\caption{Background efficiency versus shifting factor. The top plot
shows results from ANN with testing samples set 2. 
The bottom plot shows results from BDT. Dots, boxes and triangles are for
30\%, 50\% and 70\% signal efficiency, respectively.}
\end{figure}

\begin{figure}
\epsfig{figure=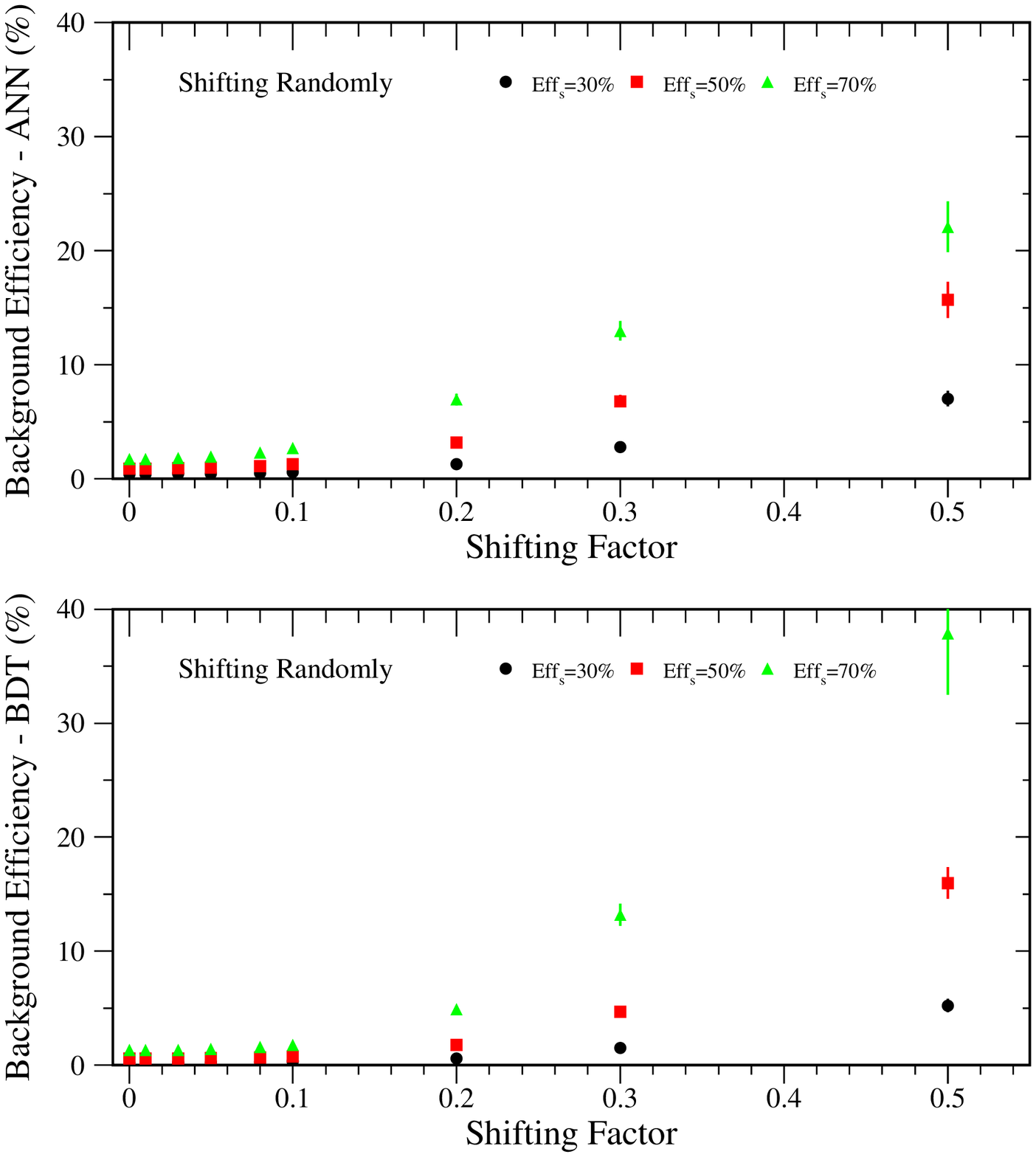,width=14cm}
\caption{Background efficiency versus shifting factor. The top plot
shows results from ANN with testing samples set 2.
The bottom plot shows results from BDT. Dots, boxes and triangles are for
30\%, 50\% and 70\% signal efficiency, respectively.}
\end{figure}


\begin{figure}
\epsfig{figure=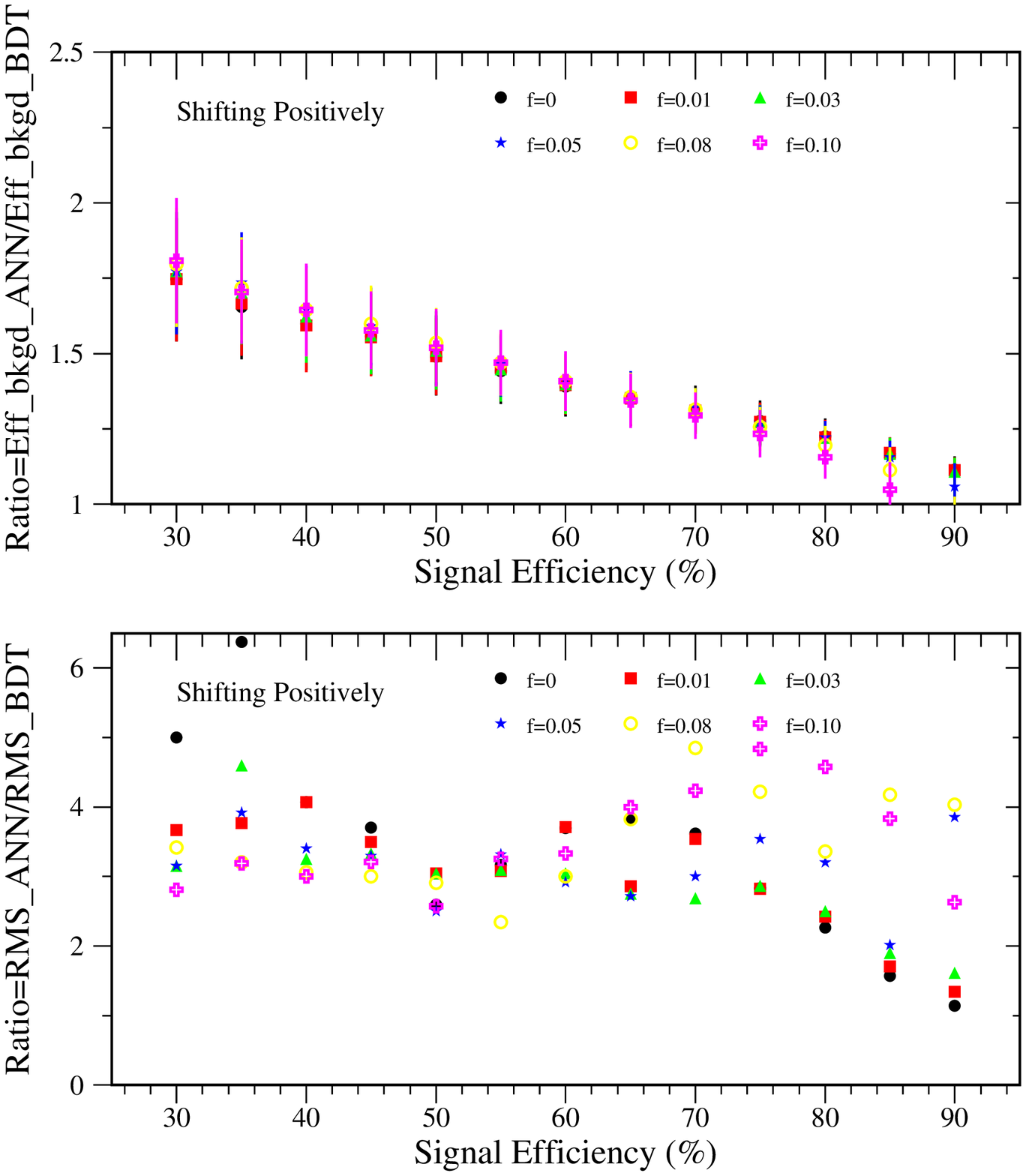,width=14cm}
\caption{Ratio of background efficiency from ANN divided by that from BDT
 versus signal efficiency(top plot) and ratio of variance from ANN divided by that
from BDT versus signal efficiency(bottom plot) with testing samples set 3.
Dots are for the testing sample without shifting; boxes, triangles, stars,
circles and crosses are for 1\%, 3\%, 5\%, 8\% and 10\% shifting, respectively.}
\end{figure}

\begin{figure}
\epsfig{figure=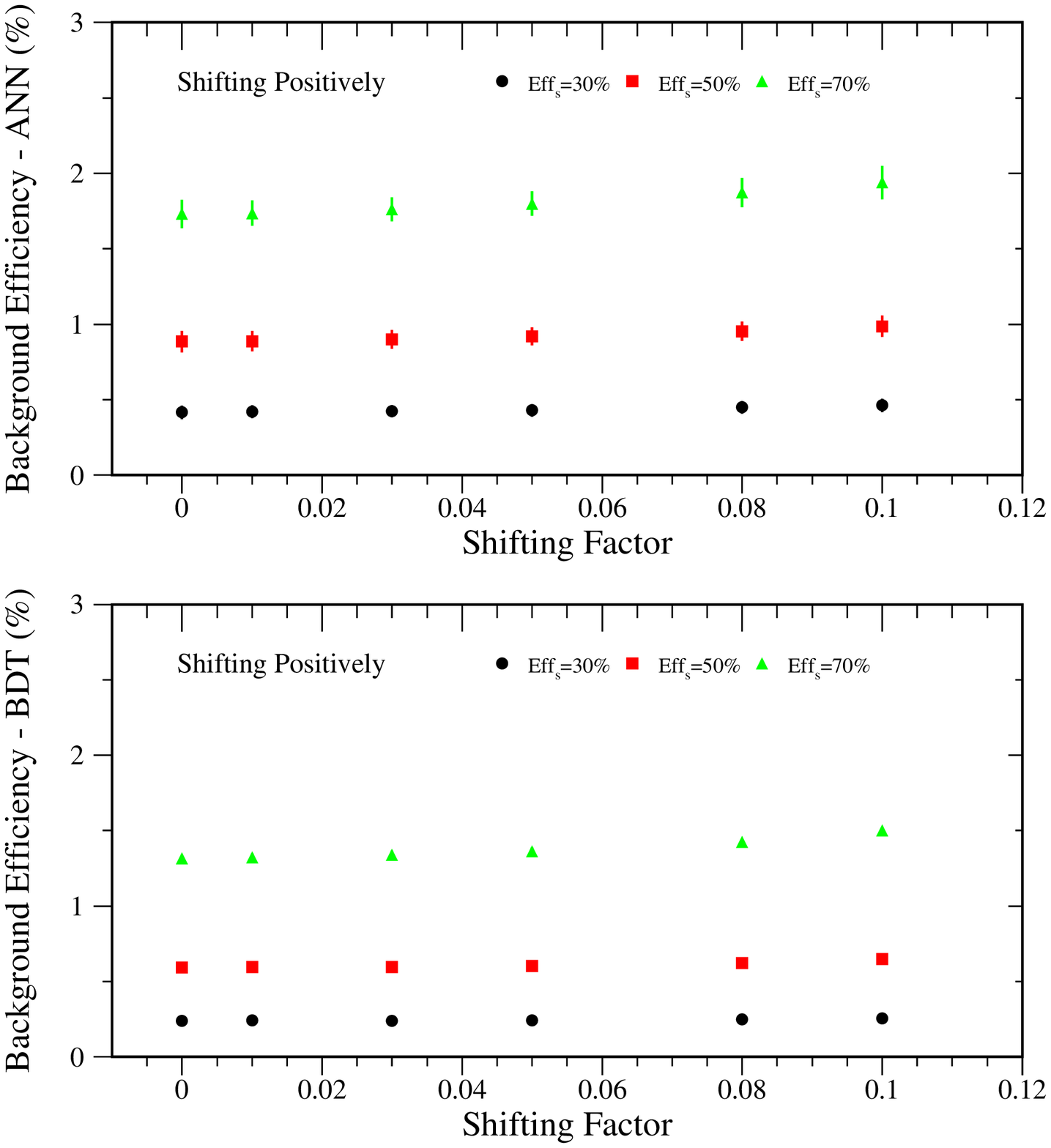,width=14cm}
\caption{Background efficiency versus shifting factor. The top plot
shows results from ANN with testing samples set 3. 
The bottom plot shows results from BDT. Dots, boxes and triangles are for
30\%, 50\% and 70\% signal efficiency, respectively.}
\end{figure}


\begin{figure}
\epsfig{figure=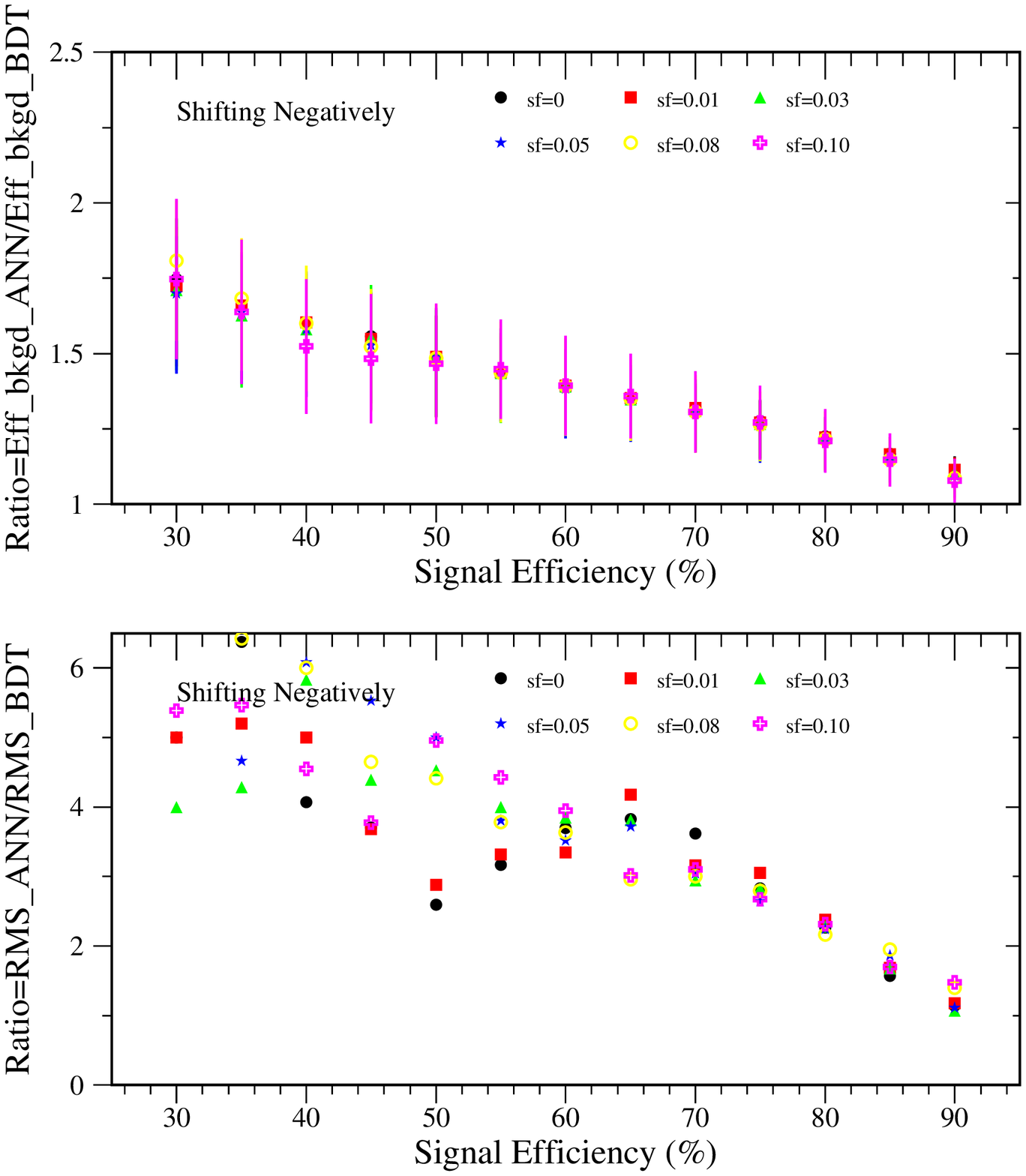,width=14cm}
\caption{Ratio of background efficiency from ANN divided by that from BDT
 versus signal efficiency(top plot) and ratio of variance from ANN divided by that
from BDT versus signal efficiency(bottom plot) with testing samples set 4.
Dots are for the testing sample without shifting; boxes, triangles, stars,
circles and crosses are for 1\%, 3\%, 5\%, 8\% and 10\% shifting, respectively.}
\end{figure}

\begin{figure}
\epsfig{figure=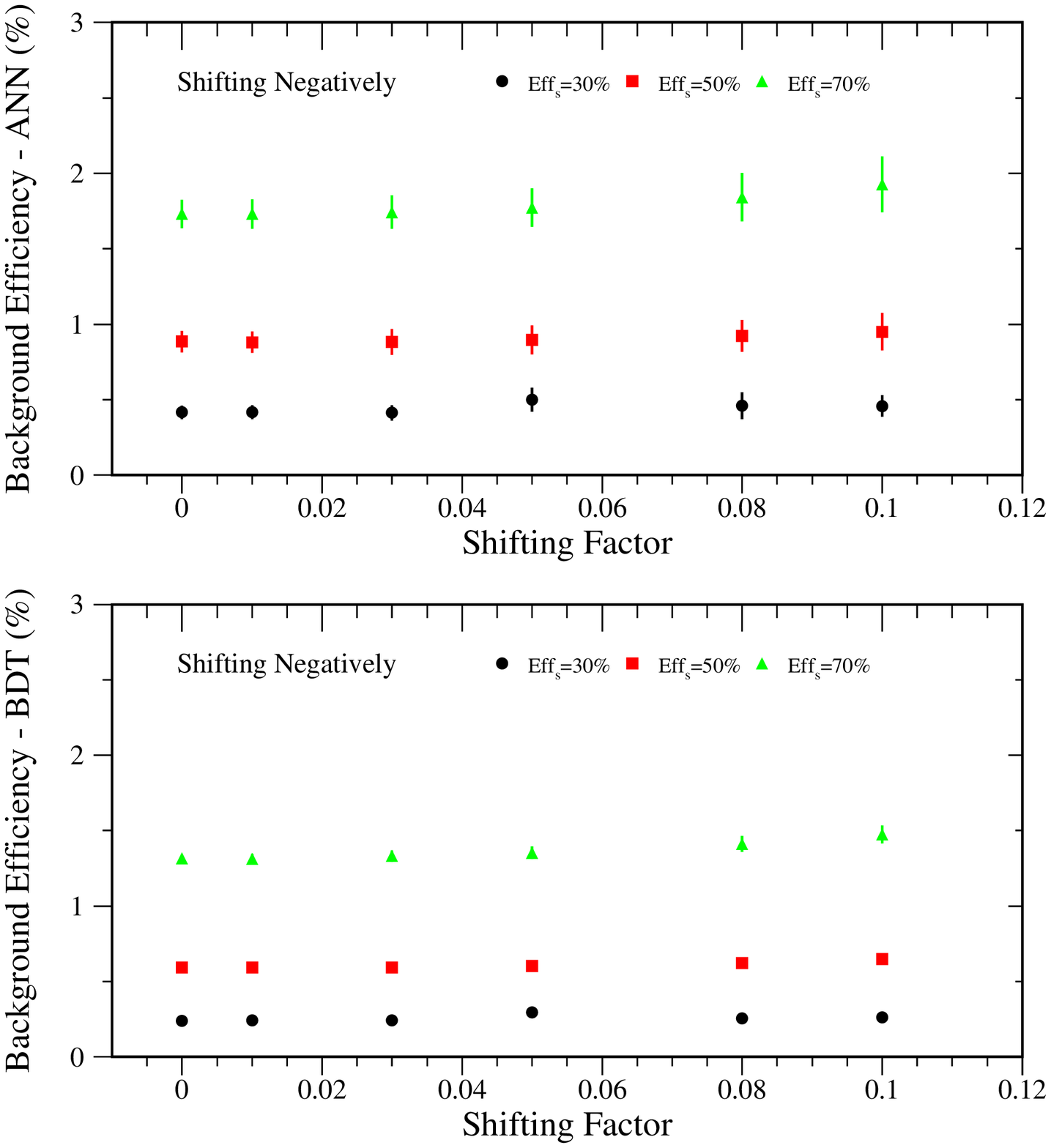,width=14cm}
\caption{Background efficiency versus shifting factor. The top plot
shows results from ANN with testing samples set 4. 
The bottom plot shows results from BDT. Dots, boxes and triangles are for
30\%, 50\% and 70\% signal efficiency, respectively.}
\end{figure}


\begin{figure}
\epsfig{figure=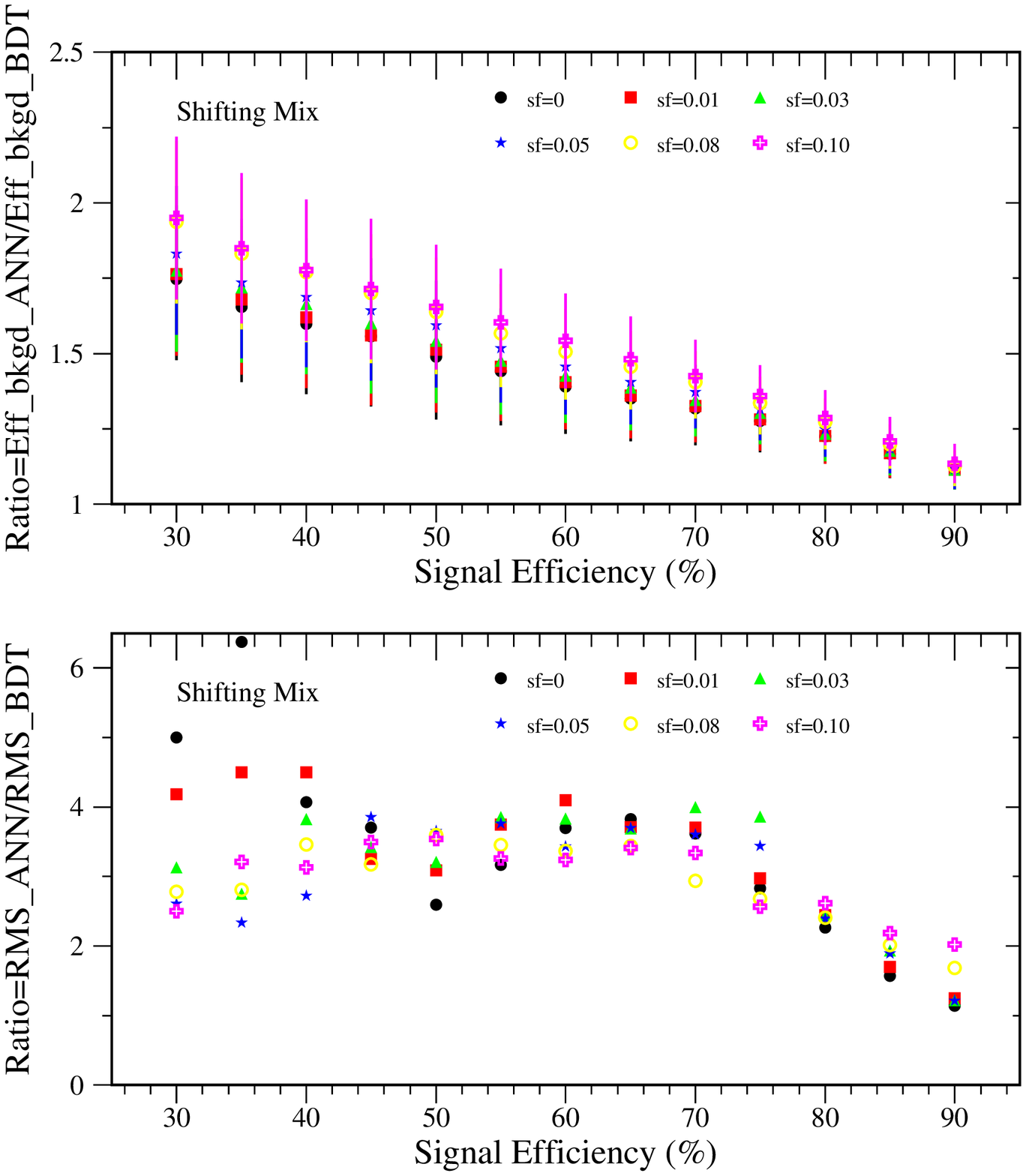,width=14cm}
\caption{Ratio of background efficiency from ANN divided by that from BDT
 versus signal efficiency(top plot) and ratio of variance from ANN divided by that
from BDT versus signal efficiency(bottom plot) with testing samples set 5.
Dots are for the testing sample without shifting; boxes, triangles, stars,
circles and crosses are for 1\%, 3\%, 5\%, 8\% and 10\% shifting, respectively.}
\end{figure}

\begin{figure}
\epsfig{figure=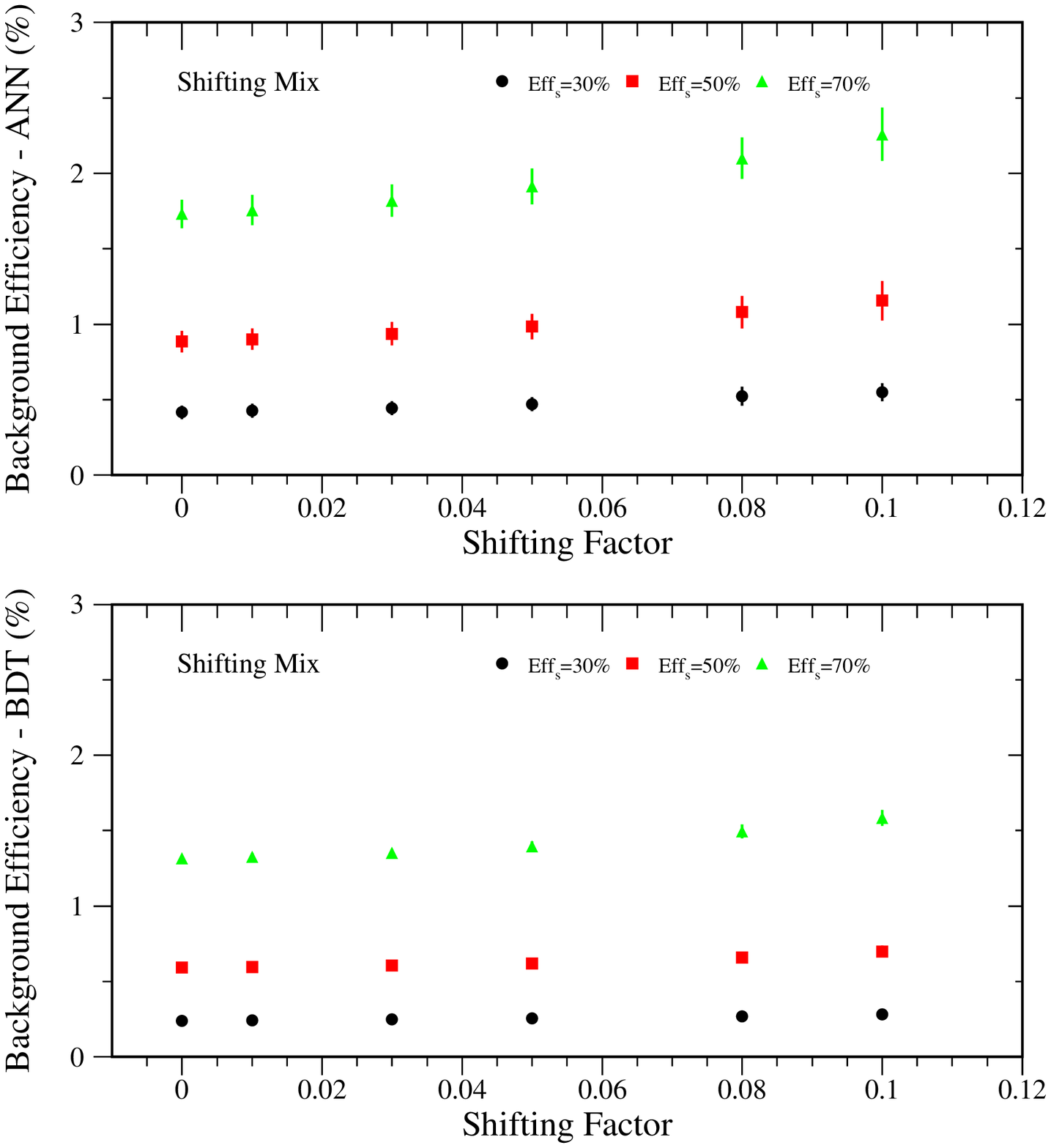,width=14cm}
\caption{Background efficiency versus shifting factor. The top plot
shows results from ANN with testing samples set 5. 
The bottom plot shows results from BDT. Dots, boxes and triangles are for
30\%, 50\% and 70\% signal efficiency, respectively.}
\end{figure}

\end{document}